\documentclass[preprint,12pt,a4paper]{elsarticle}

\usepackage{amssymb}
\usepackage{amsmath}
\usepackage{hyperref}
\DeclareGraphicsExtensions{.pdf}
\usepackage{listings}

\usepackage{color}

\definecolor{mydarkgray}{rgb}{0.3,0.3,0.3}
\definecolor{myblue}{rgb}{0.0,0.0,1.0}
\definecolor{myred}{rgb}{1.0,0.0,0.0}
\definecolor{mygreen}{rgb}{0.0,0.5,0.0}
\definecolor{mylightgray}{rgb}{0.0,0.9,0.9}

\newcommand{\brackets}[1]{\left(#1\right)}
\newcommand{\abs}[1]{\lvert#1\rvert}

\newcommand{\pd}[2]{\frac{\partial#1}{\partial#2}}

\renewcommand{\vec}[1]{\boldsymbol#1}

\newcommand{\dotp}[2]{#1\cdot#2}

\newcommand{\equ}[2]{
	\begin{equation}
	#1
	\label{#2}
	\end{equation}
}

\journal{SoftwareX}

\begin{document}
\renewcommand{\labelenumii}{\arabic{enumi}.\arabic{enumii}}

\begin{frontmatter}
%
 


\title{graph\_framework: A Domain Specific Compiler for Building Physics Applications\tnoteref{ornlfoot}}
\tnotetext[ornlfoot]{
Notice of Copyright This manuscript has been authored by UT-Battelle, LLC under Contract No. DE-AC05-00OR22725 with the U.S. Department of Energy. 
The United States Government retains and the publisher, by accepting the article for publication, acknowledges that the United States Government retains a non-exclusive, paid-up, irrevocable, world-wide license to publish or reproduce the published form of this manuscript, or allow others to do so, for United States Government purposes. 
The Department of Energy will provide public access to these results of federally sponsored research in accordance with the DOE Public Access Plan (\url{http://energy.gov/downloads/doe-public-access-plan}).
}

\author[ornl]{M. Cianciosa}
\author[didit]{D. Batchelor}
\author[ornl]{W. Elwasif}
\address[ornl]{Oak Ridge National Laboratory, PO Box 2008, MS6305 Oak Ridge TN 37831-6305, cianciosamr@ornl.gov}
\address[didit]{Diditco, Oak Ridge TN 37831 }

\begin{abstract}
Modern supercomputers are increasingly relying on Graphic Processing Units (GPUs) and other accelerators to achieve exa-scale performance at reasonable energy usage.
The challenge of exploiting these accelerators is the incompatibility between different vendors.
A scientific code written using CUDA will not operate on a AMD gpu.
Frameworks that can abstract the physics from the accelerator kernel code are needed to exploit the current and future hardware.
In the world of machine learning, several auto differentiation frameworks have been developed that have the promise of abstracting the math from the compute hardware.
However in practice, these framework often lag in supporting non-CUDA platforms.
Their reliance on python makes them challenging to embed within non python based applications.
In this paper we present the development of a graph computation framework which compiles physics equations to optimized kernel code for the central processing unit (CPUs), Apple GPUs, and NVidia GPUs.
The utility of this framework will be demonstrated for a Radio Frequency (RF) ray tracing problems in fusion energy.
\end{abstract}


\end{frontmatter}


\section*{Metadata}
\label{}

\begin{table}[!h]
\begin{tabular}{|l|p{6.5cm}|p{6.5cm}|}
\hline
\textbf{Nr.} & \textbf{Code metadata description} & \textbf{Metadata} \\
\hline
C1 & Current code version & d73b1d3 \\
\hline
C2 & Permanent link to code/repository used for this code version & \url{https://github.com/ORNL-Fusion/graph_framework} \\
\hline
C3  & Permanent link to Reproducible Capsule & \\
\hline
C4 & Legal Code License   & MIT License \\
\hline
C5 & Code versioning system used & git \\
\hline
C6 & Software code languages, tools, and services used & C++, python, Mathematica, Cuda, Metal \\
\hline
C7 & Compilation requirements, operating environments \& dependencies & cmake, netcdf, X-Code \\
\hline
C8 & If available Link to developer documentation/manual & \\
\hline
C9 & Support email for questions & cianciosamr@ornl.gov \\
\hline
\end{tabular}
\caption{Code metadata (mandatory)}
\label{codeMetadata} 
\end{table}


\section{Motivation and significance}
Standardized programming languages such as Fortran\cite{4038201}, C\cite{ritchie1993development}, and C++\cite{stroustrup2013c++}, have simplified the development of cross platform programs.
Scientific codes have relied on the ability write source code which can operate on multiple processor architectures and operating systems (OSs) with no or little changes given an appropriate compiler.
However, modern supper computers rely on graphical processing units (GPUs) to achieve exa-scale performance\cite{yang2020accelerate,8317994,9676353} with reasonable energy usage.
Unlike central processing units (CPUs), the instruction sets of GPUs are proprietary information.
Additionally, since accelerators typically are hardware accessories, an OS requires device drivers which are also proprietary.

As a consequence, attempts at standardizing cross platform programming environments such as OpenCL\cite{munshi2011opencl} and OpenACC\cite{farber2016parallel} have needed to rely on buy in from vendors which often does not materialize.
The video game industry works around this by abstracting game objects from the back-end rendering code.
Scientific codes often don't have the personnel, time, and knowledge to build these abstractions and properly exploit multi-platform GPUs.
A frameworks that can compile math equations to vendor agnostic GPU kernels would enable the scientific community to make use the new generation of exa-scale super computers.

\subsection{Compilers}
A compiler is a computer program which translates between one language and to another\cite{aho1986compilers}. 
A typical structure of a compiler translates between source code to an intermediate representation (IR)\cite{merrill2003generic,1281665}.
Optimization is performed on the IR then a back end translates the code into machine readable code.
The IR is composed of a tree structure where analysis of a program can be performed and optimizations can be performed.
Typically optimization involves simplification which reduces complexity of the tree.
Translation to machine code involves traversing this tree and outputting the instruction or instructions for each node.

\subsubsection{Auto Differentiation}
A tree or graph representation can also be used to implement automatic differentiation.
In resent years, several machine learning frameworks such as Tensorflow\cite{tensorflow2015-whitepaper} and Pytorch\cite{Ketkar2021} have emerged which use back propagation to efficiently compute gradients for training of neural networks.
However, these frameworks present several challenges for scientific codes.
The black box nature of these frameworks can result in non-optimal performance or excessive memory usage especially when computing Nth level derivatives.
Additionally their reliance on python for a front end makes them challenging to embed within C/C++/Fortran-based code for in memory model coupling.


\section{Software description}
The core of this framework is written using C++20 features.
Support for different precision levels has handled by templates.
The code is broken up into name spaces for building the computation graph, evaluating the nodes, JIT computation, workflow management, and support utilities.
Building an application consists of building an expression graph, JIT compiling kernels, and deploying the kernels in a workflow.

\subsection{\label{sub:sec:graph}Graph}
The foundation of this code is structured around a tree data structure that enables the symbolic evaluation of mathematical expressions.
The \texttt{graph} name space contains classes which symbolically represent mathematical operations.
Each node of the graph is defined as class derived from a \texttt{leaf\_node} base class.
The \texttt{leaf\_node} defines virtual methods to \texttt{evaluate}, \texttt{reduce}, \texttt{df}, \texttt{compile}, and methods to support introspection.
A feature unique to this code compared to other graph frameworks is the ability to render the expression trees to \LaTeX which aids debugging.

Sub-classes of \texttt{leaf\_node} include end nodes for constants, variables, arithmetic, basic math functions, and trigonometry functions.
Other nodes encapsulate more complex expressions like piece wise constants which depend on the evaluation of an argument.
These piece wise constants are used implement spline interpolation expressions.

Each node is constructed via factory methods.
For common arithmetic operations, the framework overloads the $+-*/$ operators to construct expression nodes.
The factory method checks a \texttt{node\_cache} to avoid building duplicate sub-graphs.
Identification of duplicate graphs is performed by computing a hash of the sub-graph.
This hash can be rapidly checked if the same hash already exists in a \texttt{std::map} container.
If the sub-graph already exists, the existing graph is returned otherwise a new sub-graph is registered in the \texttt{node\_cache}.

Each time an expression is built, the \texttt{reduce} method is called to simplify the graph.
For instance, a graph consisting of constant added to a constant will be reduced to a single constant by calling the \texttt{evaluate} method.
Sub-graph expressions are combined, factored out, or moved to enable better reductions on subsequent passes.
As new ways of reducing the graph are implemented, current and existing code built using this framework benefit from improved speed.

Complex mathematical expressions are defined by chaining nodes together.
Auto differentiation is handled by implementing the chain rule for a node using the \texttt{df} method.
This method builds new tree expressions by applying the chain rule with respect to a different \texttt{leaf\_node}.
At the ends of the expression tree, derivatives of \texttt{constant} nodes return zero while derivatives \texttt{variables} return one or zero depending if the derivative is taken in respect to itself.

\subsection{\label{sub:sec:jit}JIT}
Once expression trees are constructed, these can be JIT compiled to a back-end.
The graph framework supports back ends for generic CPUs, Apple Metal GPUs, Nvidia Cuda GPUs, and initial HIP support of AMD GPUs.
Each back end supplies relevant driver code to build the kernel source, compile the kernel, build device data buffers, and handle data synchronization between the device and host.
All JIT operations are hidden behind a generic \texttt{context} interface.
Kernel source code is built by defining the kernel input variables, output nodes, and setter maps.
Setter maps enable variable updates such as time stepping by mapping a graph output back to a node input.

Each context, creates a specific kernel preamble and post-fix to build the correct syntax.
Memory access is controlled by loading memory once in the beginning, and storing the results once at the end.
Kernel source code is built by recursively traversing the graph and calling the \texttt{compile} method of each \texttt{leaf\_node}.
Each line of code is stored in a unique register variable assuming infinite registers.
Duplicate code is eliminated by checking if a sub-graph has already been traversed.
Once the kernel source code is built, the kernel library is compiled, and a kernel dispatch function is created using a C++ lambda function.

For CPU back-ends, the source code is compiled in memory to LLVM-IR\cite{1281665} then JITed to machine code using the LLVM MCJIT library.
Metal, CUDA, and HIP kernels are JIT compiled using available functions in their respective application programming interfaces (APIs).
Each context includes special kernels for reduction operations.
To implement a new back end requires building a context object which encapsulates a vendors API calls.

\subsection{\label{sub:sec:workflow}Workflow}
Once the kernels are created, they can be called in sequences using the \texttt{manager} object.
A manager object encapsulates \texttt{work\_item} or \texttt{converge\_item} objects which encapsulate a  kernel call.
Using this \texttt{manager} workflows like newton solver can be implemented by chaining calls to evaluation and reduction kernels until a tolerance is met.


\section{Illustrative examples}
\begin{lstlisting}[caption={Example source code to build expression graphs, JIT compile them, and run them on a compute resource. The $*$ and $+$ operator are overloaded to build graph \texttt{multiply} and \texttt{add} nodes. A graphical representation of expression trees created are shown in Figure \ref{fig:Tree}.},captionpos=b,label={code:example},language=C++]
template<jit::float_scalar T>
void example(const size_t id) {
// Build Expression Trees
  auto x = 
    graph::variable<T> (100000, 'x');
  auto y = 10.0*x + 0.5;
  auto dydx = y->df(x);

// JIT Compile and run expressions.
  workflow::maganger<T> work(id);
  work.add_item({
    graph::variable_cast(x)
  }, {y, dydx}, {}, {}, "name");
  work.compile();
  work.run();
  work.wait();
}
\end{lstlisting}

\begin{figure}
\begin{center}
\includegraphics[width=\columnwidth]{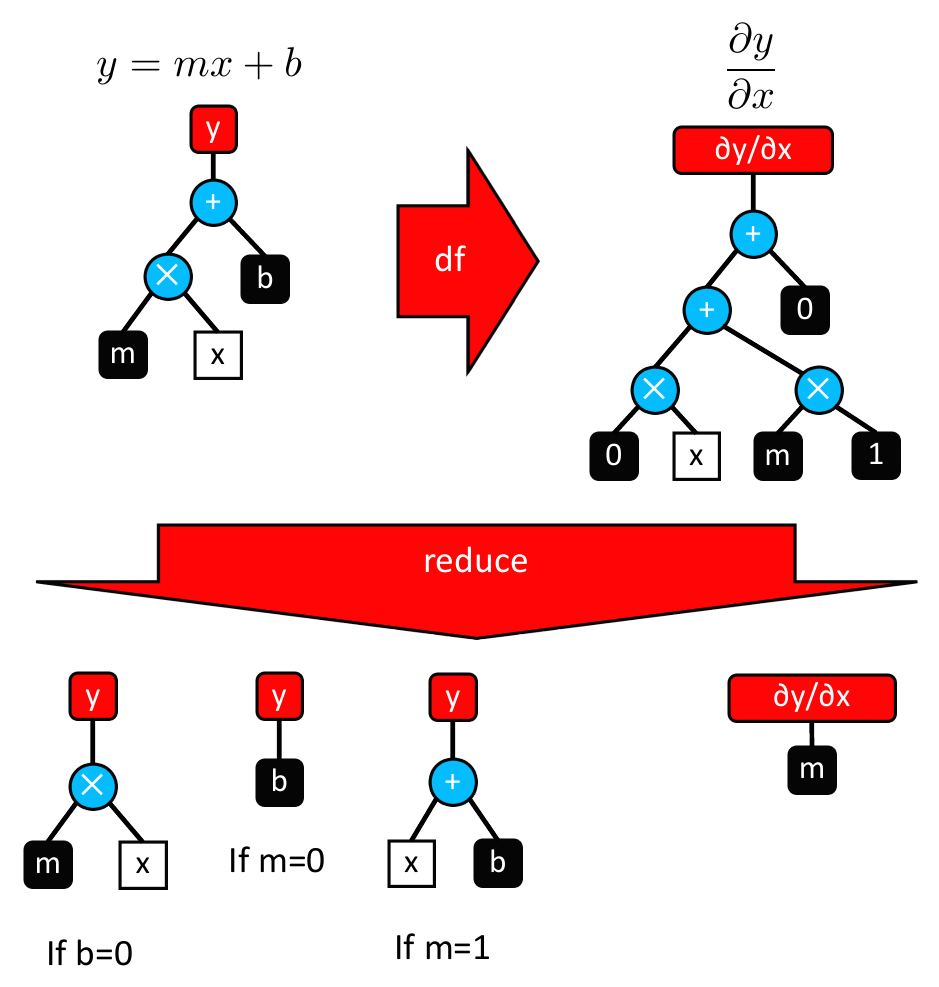}
\end{center}
\caption{Mathematical operations are defined as a tree of operations.
A \texttt{df} method transforms the tree by applying the derivative chain rule to each node.
A \texttt{reduce} method applies algebraic rules removing nodes from the graph.}
\label{fig:Tree}
\end{figure}

Example source code for building expressions is shown in Listing \ref{code:example}.
In this example, a variable is defined for $x$ then the expressions for a line and slope of a line are build from it.
Figure \ref{fig:Tree} shows a visualization of the expression three created from the source code in Listing \ref{code:example}.
An expression tree is built for $y$.
The \texttt{df} method with an argument of the $x$ variable is called on the $y$ expression.
This runs the \texttt{df} method recursively on each node of the tree creating a new expression for $\pd{y}{x}$.
The \texttt{reduce} method finds opportunities to eliminate parts of the graph.
For addition nodes, a zero in either right or left branch eliminates the need for the addition operation.
For multiplication nodes, a one or a zero in either branch eliminates the multiplication node.
Combining these two methods results in the simplest expression for $\pd{y}{x}$.
Note in the actual framework, the full expression tree for $\pd{y}{x}$ is never created since the \texttt{reduce} method is reducing the graph on the fly as each new node is created.

The expression trees are compiled into kernels and a kernels are arranged into workflows.
First a workflow manager object is created for the current thread.
A work item is added to the manager to evaluate the graph expressions built on lines 5 and 6 of Listing \ref{code:example}.
This will build a kernel with one input $x$ and two outputs $y$ and $dydx$ called \texttt{name}.
The \texttt{compile} method JIT compiles all the kernels then the \texttt{run} method runs all the kernels in the order they were created.
Since GPUs operate asynchronously, we need to explicitly wait for kernels to finish.


\section{Impact}

\begin{figure}
\begin{center}
\includegraphics[width=0.9\columnwidth]{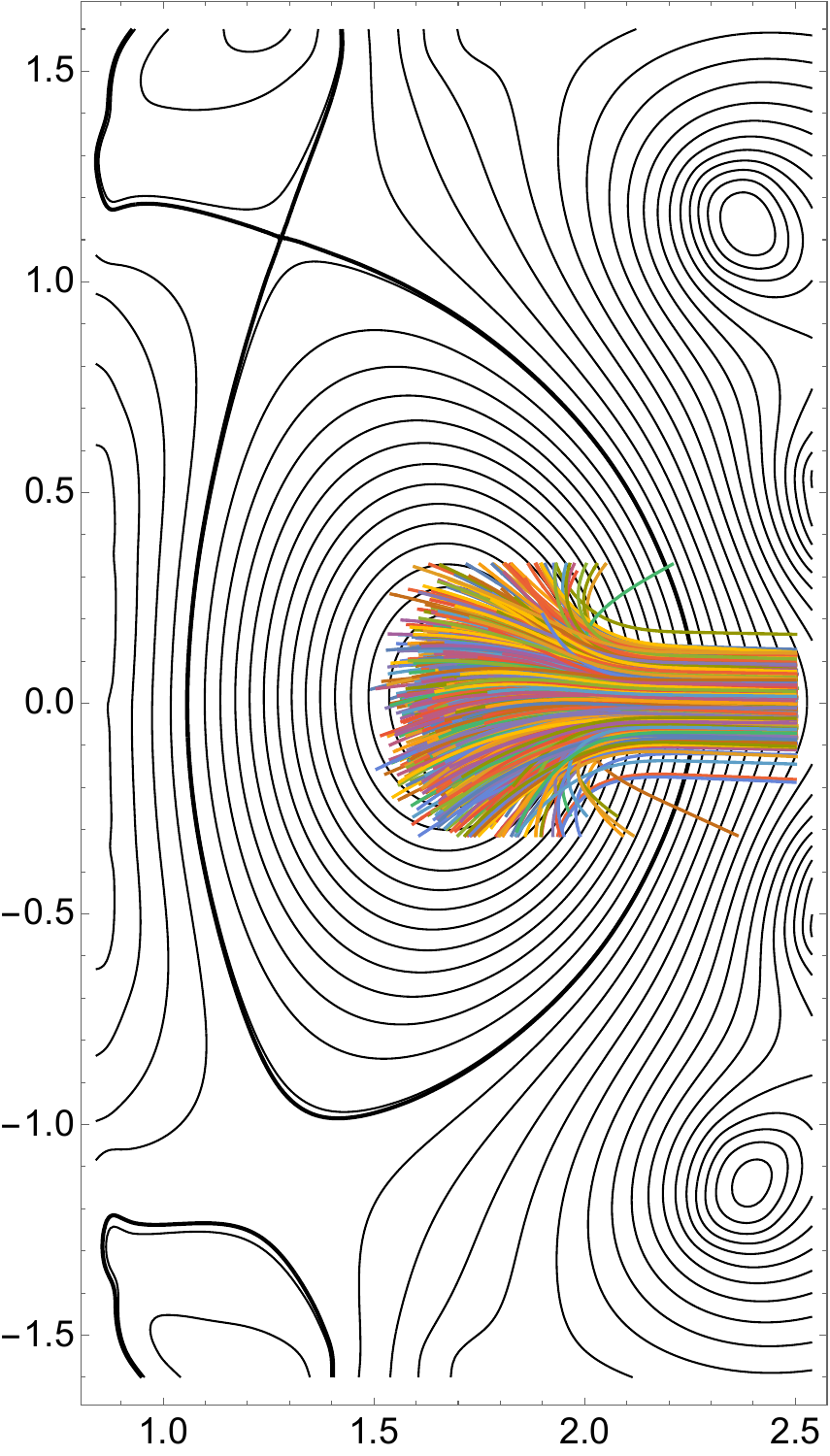}
\end{center}
\caption{Ray trajectory for $1\times10^{5}$ rays traced in a realistic tokamak geometry.}
\label{fig:TokamakRays}
\end{figure}

There are many problems in fusion energy where the same physics needs to be applied to a large ensemble.
Understanding the impacts of runaway electron populations\cite{10.1063/1.4981209,10.1063/5.0022072}.
Generating large data sets of RF heating efficiency to build reduced models\cite{Irvin25042025}.
In some of these domains the disparate time scales necessitates computational efficiency\cite{HOPPE2021108098}.
Understanding full orbit affects on particle losses in the edge\cite{deGrassie_2009} and how they impact the first wall.
The framework described here lowers the barriers for domain physicists to efficiently utilize GPU resources.
Using the framework described in the previous sections, a modern GPU capable RF Ray Tracing code was developed.
The abstractions afforded by this framework allow arbitrary geometry and easy extension to new physics.

Geometric optics is a set of asymptotic approximation methods to solve wave equations.
The physics of the particular wave determines an algebraic relation between $\omega$ and $\vec{k}$ called a dispersion relation, $D\brackets{\omega,\vec{k}}=0$.
Since the parameter $t$ does not appear explicitly in the dispersion relation, the function $\omega{\brackets{\vec{k}\brackets{t},\vec{x}\brackets{t}}}$ is constant along the ray trajectory
\equ{
\pd{\omega}{t}=\dotp{\pd{\omega}{\vec{x}}}{\pd{\vec{x}}{t}}+\dotp{\pd{\omega}{\vec{k}}}{\pd{\vec{k}}{t}}\equiv 0
}{equ:ray_w}
by virtue of the ray equations.
Since the dispersion relation is satisfied all along the ray trajectory, the derivatives needed for the ray equations can be obtained by implicit differentiation
\equ{
\pd{D}{\vec{x}}=\pd{D}{\omega}\pd{\omega}{\vec{x}}\Rightarrow\pd{\omega}{\vec{x}}=-\frac{\pd{D}{\vec{x}}}{\pd{D}{\omega}}
}{equ:implicit_x}
\equ{
\pd{D}{\vec{k}}=\pd{D}{\omega}\pd{\omega}{\vec{k}}\Rightarrow\pd{\omega}{\vec{k}}=-\frac{\pd{D}{\vec{k}}}{\pd{D}{\omega}}
}{equ:implicit_k}
These are the equations that are actually integrated.

A ray tracing problem is build by implementing expressions for the plasma equilibrium and a dispersion relation.
Equations of motion are defined using the auto differentiation.
Expressions for ray update are constructed using the expressions of 4th order Runga-Kutta.
These expressions are JIT compiled into a single kernel call with inputs for $\vec{x}$, $\vec{k}$, $t$, and $\omega$ with outputs for the dispersion residual, and step updates for $\vec{x}$, $\vec{k}$ and $t$.
Figure \ref{fig:TokamakRays} shows $1\times10^{5}$ O-Mode rays traced in a realistic tokamak geometry.

In a spatially varying medium, at a given frequency, there may be regions in which the solution of the dispersion relation, $\vec{k}$, is real, and the wave propagates.
In other regions $\vec{k}$ is imaginary and the wave does not propagate, referred to as evanescent.
The boundary between a region of propagation and evanescence is a surface called a cut-off.
It is also possible that surfaces occur where $\vec{k}$ diverges to infinity, in which case the phase velocity component normal to the surface goes to zero.
These are called resonances.
Typically, the wave is reflected at a cut-off and is absorbed or converted to a different type of wave at a resonance. 
These critical surfaces, therefore, denote important changes in wave behavior, and the behavior of rays in their vicinity is an indication of the correctness of the solution.

For plasmas, the spatial dependence of the dispersion relation comes through variation of the plasma equilibrium quantities.
These include the vector magnetic field, $\vec{B}\brackets{x}$, the density of each plasma particle species, $n_{s}\brackets{x}$, and the temperature of each particle species, $T_{s}\brackets{x}$, where $s$ indicates a particular species.
For the cases presented here a linear gradient along the $x$ direction is taken for either the particle density or magnetic field strength.
\equ{
f\brackets{x}=0.1x+1
}{equ:grad}

Initial conditions for the wave solution are obtained by choosing fixed values for $\omega$, $\vec{x}$, $k_{y}$, and $k_{z}$.
The remaining value $k_{x}$ is determined using a Newton method to a tolerance of $\abs{D\brackets{\vec{k},\omega}}<10^{-15}$.
Since the dispersion functions are multi-valued, an initial guess for $k_{x}$ selects among the possible roots. 


\subsection{Cold Plasma Dispersion Relation}

\begin{figure}
\begin{center}
\includegraphics[width=0.8\columnwidth]{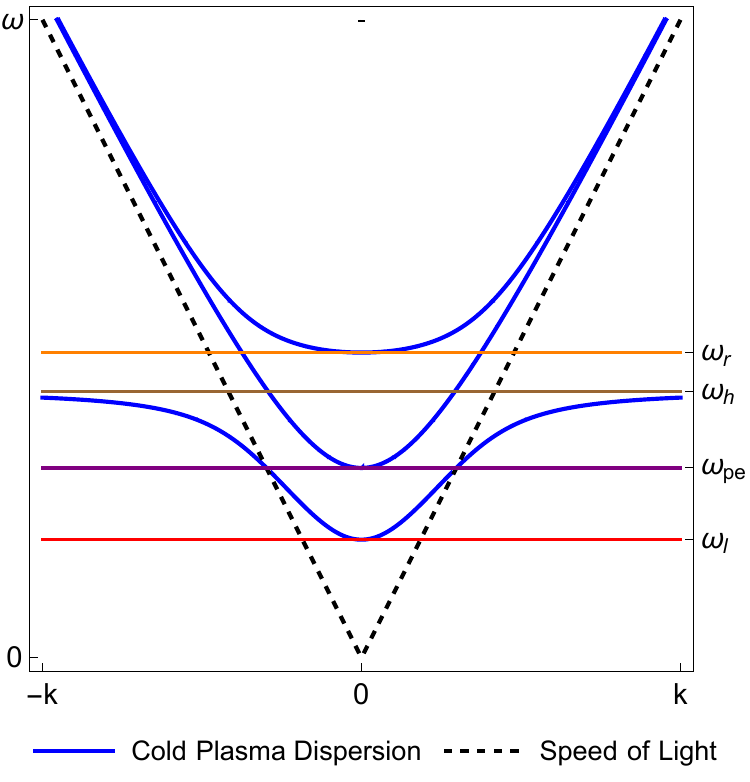}
\end{center}
\caption{The O-Mode branch can propagate through the quiescent region between the right-hand cutoff, $\omega_{r}$, and the upper hybrid resonance, $\omega_{h}$, that the X-Mode branch cannot but is cut off at the plasma frequency, $\omega_{pe}$.
The X-Mode branch can pass the O-Mode's plasma cutoff but is stopped by the left-hand cutoff, $\omega_{l}$.}
\label{fig:cold-plasma-disperion}
\end{figure}

\begin{figure}
\begin{center}
\includegraphics[width=\columnwidth]{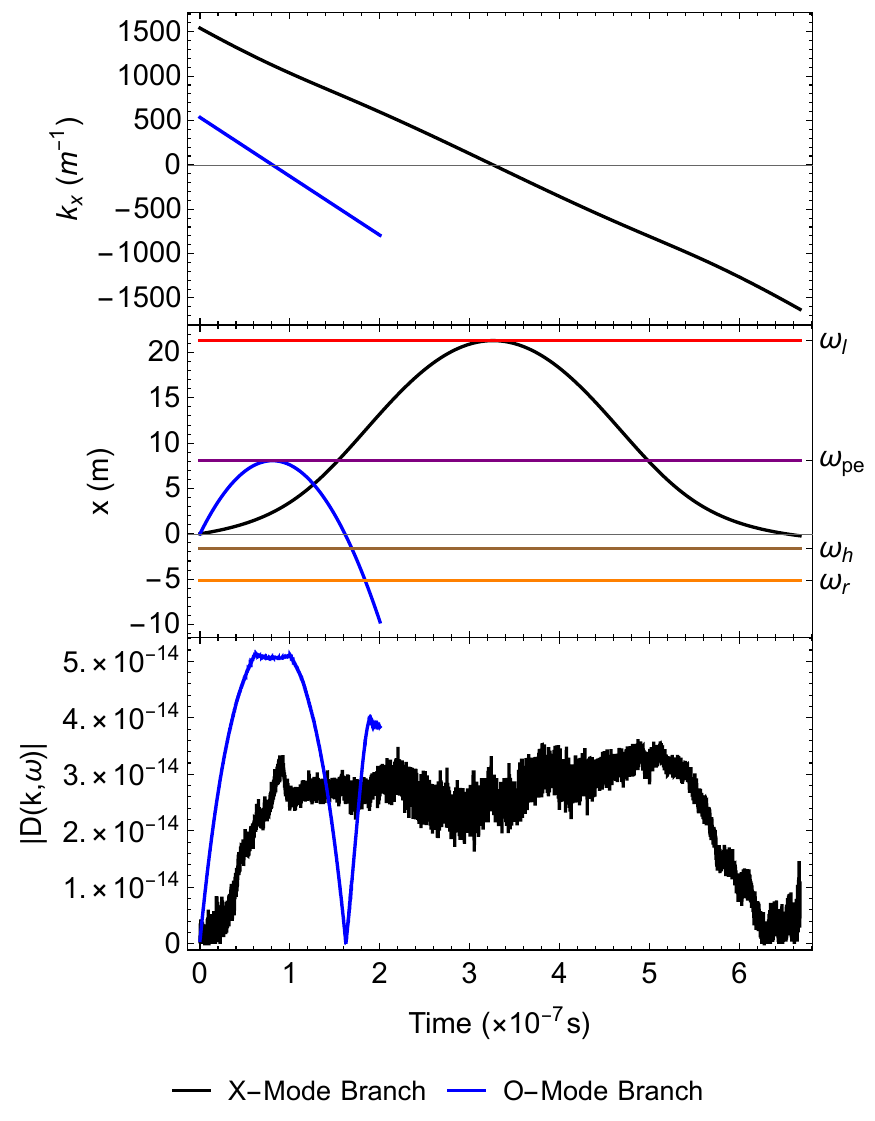}
\end{center}
\caption{Wave trajectories with frequencies between the upper hybrid resonance, $\omega_{h}$, and the left-hand cutoff, $\omega_{l}$, for the cold plasma dispersion relation.
The X-Mode branch can pass the plasma frequency cutoff, $\omega_{pe}$, while O-mode cannot.
The O-Mode can pass through the upper hybrid resonance, $\omega_{h}$, while the X-Mode branch is absorbed.
The bottom plot tracks the resulting dispersion function residual, $\abs{D\brackets{k,\omega}}$, of the solver as the rays are traced.}
\label{fig:cold-plasma-result2}
\end{figure}

The general cold plasma dispersion relation, valid for plasmas with multiple particle species in the cold plasma limit and for arbitrary frequency, is of the form 
\equ{
D\brackets{\vec{k},\omega}=Det\brackets{\vec{\epsilon}+\vec{n}\vec{n}-n^{2}\vec{I}}
}{equ:cold_plasma}
where
\equ{
\vec{n}=\vec{n}_{||}+\vec{n}_{\perp}=\brackets{\vec{k}_{||}+\vec{k}_{\perp}}\frac{c}{\omega}
}{equ:cold_plasma_n}
and $\vec{\epsilon}$ is the dielectric tensor.
Using the Onsager symmetries, this tensor is defined as
\equ{
\vec{\epsilon}=\left(
\begin{array}{ccc}
\epsilon_{11}  & \epsilon_{12} & 0             \\
-\epsilon_{12} & \epsilon_{11} & 0             \\
0              & 0             & \epsilon_{33}
\end{array}
\right)
}{equ:cold_plasma_e}
for a cold plasma.
The elements of this tensor are
\equ{
\epsilon_{11}=1-\sum_{s}\frac{\frac{\omega^{2}_{p}}{\omega^{2}}}{1-\frac{\Omega^{2}_{c}}{\omega^{2}}}
}{equ:cold_plasma_e11}
\equ{
\epsilon_{12}=-i\sum_{s}\frac{\frac{\Omega_{c}}{\omega}\frac{\omega^{2}_{p}}{\omega^{2}}}{1-\frac{\Omega^{2}_{c}}{\omega^{2}}}
}{equ:cold_plasma_e12}
\equ{
\epsilon_{33}=1-\sum_{s}\frac{\omega^{2}_{p}}{\omega^{2}}
}{equ:cold_plasma_e33}
where $\omega_{p}$ is the plasma frequency and $\Omega_{c}$ is the cyclotron frequency for a species $s$.

Figure \ref{fig:cold-plasma-disperion} shows the dispersion relation for a uniform magnetic field and density gradient.
This dispersion is a superposition of the O-Mode and X-mode dispersion relations.
For a given frequency, one branch can cross cutoffs and resonances, the other cannot.
Figure \ref{fig:cold-plasma-result2} shows the wave trajectories for two waves between $\omega_{h}$ and $\omega_{l}$.
The X-Mode branch can pass $\omega_{pe}$ cutoff while the O-Mode branch is reflected.
The X-Mode branch is absorbed in the upper hybrid resonance, $\omega_{h}$, while the O-Mode branch can pass through it.

\subsection{Performance Scaling}

\begin{figure}
\begin{center}
\includegraphics[width=\columnwidth]{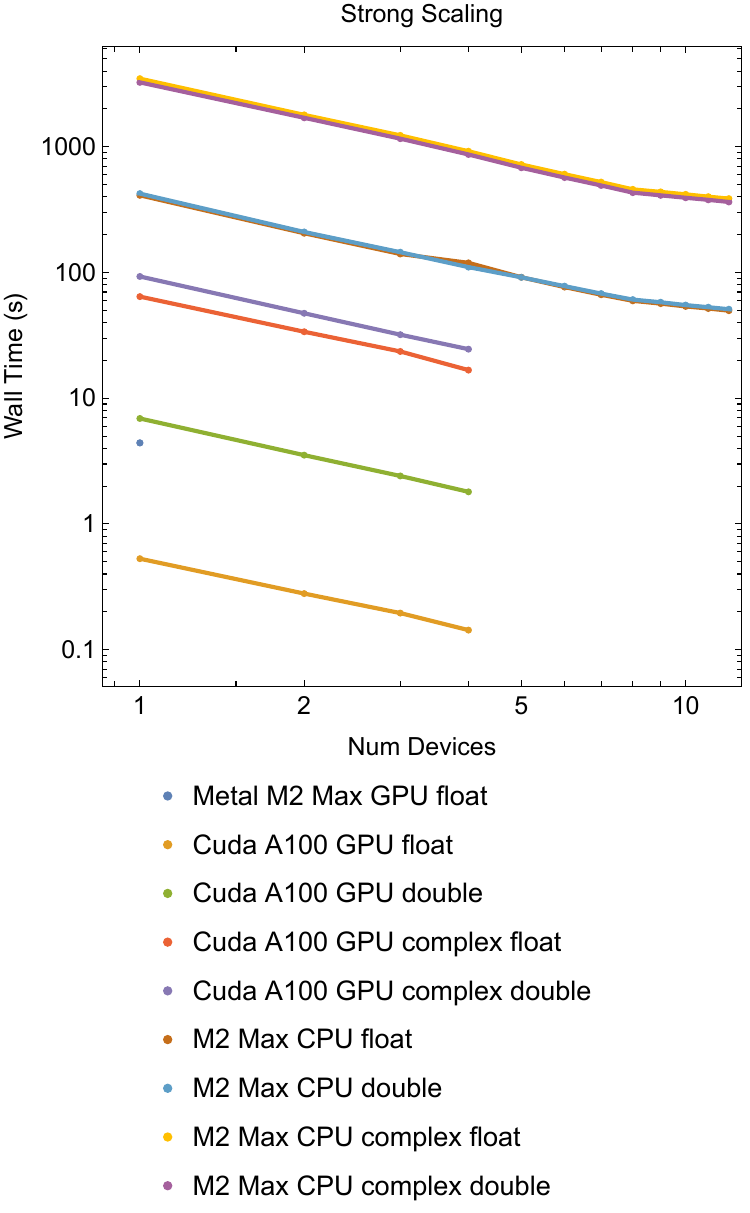}
\end{center}
\caption{Wall time strong scaling for 1E6 Rays traced in a realistic tokamak equilibrium. 
Benchmarking was performed on a Mac Studio with an 12 core M2 Max with a single GPU.
The other machine features up to 4 Nvidia A100 GPUs.}
\label{fig:benchwalltime}
\end{figure}

To benchmark code performance we traced $1\times 10^{6}$ rays using the cold plasma dispersion relation in a realistic tokamak equilibrium. 
Figure \ref{fig:benchwalltime} shows the strong scaling of wall time as the number of GPU and CPU devices are increased.
Bench marking was prepared on two different setups.
The first set up was a Mac Studio with an Apple M2 chip.
The M2 chip contains a 12 core CPU where 8 cores are faster performance codes and the remaining 4 are slower efficiency cores.
An interesting thing to note is the Apple M series CPU processors show almost no performance difference between single and double precision.
The M2 also contains a single 38-core GPU that only support single precision operations.
The second setup is a server with 4 Nvidia A100 GPUs.
Bench marking measures the time to trace the $1\times10^{6}$ rays but does not include the setup and JIT times.


\section{Conclusions}
Building a graph data structure representation of equations enables a very powerful tool for compiling physics expressions to optimized kernels.
A graph form enables auto differentiation and symbolic manipulation of mathematical expressions.
As an example in the Ray tracing problem, new dispersion relations can be implemented without regard to equilibrium geometry.
Auto differentiation automatically produces gradient terms.
Symbolic reduction can automatically remove terms based on symmetries in the problem.
By contrast in a legacy code, either the expressions would need to assume a specific symmetries or gradient terms would be to be explicitly defined.

These symbolic expression trees can be used to generate source code different CPUs and GPUs.
Bench marking shows kernels generated can efficiently scale to multiple CPU and GPU devices accross a variety of vendors.
The examples presented here show waves or particles in a fusion plasmas but the framework is applicable to any physics problem involving a large ensemble of independent problems.


\section*{Acknowledgements}
The authors would like to thank Dr. Yashika Ghai, Dr. Rhea Barnett, and Dr. David Green for their valuable insights when setting up test cases.




\bibliographystyle{elsarticle-num}
\bibliography{Rays.bib}

\end{document}